\documentclass[aps, prd, preprint, groupedaddress,12pt]{revtex4}
\usepackage{mathrsfs}
\usepackage{bbm}
\usepackage{amsfonts}
\usepackage{amsmath}
\usepackage{graphicx}
\usepackage{subfig}
\usepackage[T1]{fontenc}
\usepackage{feynmp}

\newcommand{\void}[1]{}

\def\be{\begin{equation}}
\def\ee{\end{equation}}
\def\bea{\begin{eqnarray}}
\def\eea{\end{eqnarray}}
\def\lmd{\lambda}
\def\bt{\beta}

\newcommand{\ppt}[1]{{\partial\over \partial #1}}
\newcommand{\td}[1]{\tilde{#1}}

\newcommand{\bary}{\begin{array}}
\newcommand{\eary}{\end{array}}

\def\nb{\nonumber}

\begin{document}

    \title{Recursion relations for the general tree-level amplitudes in QCD with massive dirac fields}

\author{
Gang~Chen,$^{1}$}
{\affiliation{
$^{1}$Department of Physics, Nanjing University\\
22 Hankou Road, Nanjing 210098, China
}

\hspace{1cm}
\begin{abstract}
QCD amplitudes with many external fields have been studied for a long time. At tree-level, the amplitudes can be obtained effectively by the Britto-Cachazo-Feng-Witten (BCFW) recursion relations. In this article, we extend the BCFW relations to the QCD amplitude of which the external fields are all massive or include only one massless line. We find such amplitude can be split into two parts and each part of the amplitude is of some correlated spin configuration between the two shifted lines. After choosing proper momentum shift scheme, we can show that each part is constructible directly. Hence, we can obtain a general procedure for the amplitudes in QCD by the BCFW recursion relations. We apply the procedure to several amplitudes as examples. We find such methods are very efficient when there are many massive external fields in the amplitudes.
\end{abstract}


\date{\today}
\maketitle

\section{Introduction}
Tree-level QCD amplitude have been well-study for a long time \cite{Parke:1986gb, Xu:1986xb, Berends:1987me, Dixon1, Dixon2, Dixon3, Dixon4,Chalmers}. As the analysis in \cite{Witten1}, the spinor form of perturbative gauge theory have a explain in string theory in twistor space. After that, the on-shell BCFW recursion relation was proposed in  \cite{Britto:2004nj,Britto:2004nc,Britto:2004ap,Britto:2005dg,Britto:2005fq, Britto:2005ha,Cachazo:2004dr,Cachazo:2004kj,Cachazo:2004zb,Cachazo:2004by, Cachazo:2005ca}. This provide a very efficient method for the calculation of the amplitudes in many theories \cite{Badger1,Badger2,Ozeren,Schwinn,cohen,Georgiou:2010mf,Cachazo1}. Such relation is accomplished by shifting  the external momentum $p(z)$ such that the external fields are still on shell and the momentum conservation hold. The amplitude with shifted external momentum are denoted as $\mathcal{A}(z)$. The recursion relations are effective tools if the following three conditions \cite{Britto:2005fq} are met. (1) Rational condition: $\mathcal{A}(z)$ is rational, (2) Constructibility condition: it vanish for $z\rightarrow\infty$, (3) Simple pole condition: the only singularities are simple poles.

In QCD, for tree-level amplitude with general momentum, the first and third condition are usually met. However, the constructibility condition is not always the case. In fact, for the amplitudes with all massive dirac fields or only one massless gluon, we can not choose a momentum shift scheme such that the amplitude is constructible directly. To get over this obstruction, we split the general amplitude as a linear combination of two limited amplitudes. Each of them are constructible under some choice of the momentum shift scheme. This accomplishment mainly due to linear properties of the amplitudes with respect to external fields. Hence we can also use the BCFW recursion relations to obtain the general amplitudes efficiently.

In fact such calculation can be even more simplified. The amplitude with different spin configurations for the massive lines are related. By acting with the raising or lowering operators \cite{chen} of the little group on the external fields, we can get the amplitude with all spin configurations from just one.. The three generators of the little group $SO(3)$ for massive fields can be represented as the first-order differential operators with respect to spinors \cite{chen}
\bea
\mathcal{R}(J^1)&=&{-1\over 2} \left(\bt\ppt{\lmd}-\td\bt\ppt{\td\lmd}+\lmd\ppt{\bt}-\td\lmd\ppt{\td\bt} \right),\nb\\
\mathcal{R}(J^2)&=& {i\over 2} \left(\bt\ppt{\lmd}+\td\bt\ppt{\td\lmd}-\lmd\ppt{\bt}-\td\lmd\ppt{\td\bt} \right), \nb\\
\mathcal{R}(J^3)&=& {-1\over 2} \left(\lmd\ppt{\lmd}-\td\lmd\ppt{\td\lmd}-\bt\ppt{\bt}+\td\bt\ppt{\td\bt} \right).
\eea
The conventions can be find in \cite{chen}. In the particular calculations, the raising and lowering form are more convenient
\bea\label{Jpmspinor}
\mathcal{R}(J^+)&=& \left(\td\lmd\ppt{\td\bt}-\bt\ppt{\lmd} \right)\nb\\
\mathcal{R}(J^-)&=& \left(\td\bt\ppt{\td\lmd}-\lmd\ppt{\bt}\right).
\eea

In this article, we organize as follows. In Section \ref{MDfield}, we discuss several kinds of momentum shift scheme and the corresponding $z$-independent spin states for the massive fields. In Section \ref{Sec-Gamp}, we explore the general procedures for the amplitude in QCD, especially for the case when external fields are all massive or contain only one massless gluon. We apply the procedure to some simple amplitudes and compare the results with the usual Feynman rules. Absolutely they are consistent with each other.

\section{The momentum shift scheme for the amplitudes in QCD}\label{MDfield}
If the shifted two lines are all massless, the momentum shift and the application to the amplitude are well-known \cite{Britto:2004ap,Badger1,Badger2}. In this section, we analysis the several kinds of two-line momentum shift schemes when some of  the external line are massive fields. Under each shift scheme, we find it is possible to choose a spin state such that they are z-independent for the massive fields. For the massless fields, we can also choose a shift scheme such that the external wave function are proportional to $1\over z$ when the shift parameter $z$ tend to infinity. Then we analysis the  forms of the amplitudes in this limit. Since we focus on the amplitude is QCD, the massive and the massless lines are dirac fields and gluon fields respectively.

We first discuss the momentum shift for two massive fields. We denote the shifted lines as $\hat q_1,\hat q_2$. Since the amplitudes are Lorentz invariant, we can choose a reference frame such that the two shifted momentum can be of form
\bea\label{twoLS}
p_{\hat q_1}&=&\lmd_{q_1}\td\lmd_{q_1}+\bt_{q_1} \td\bt_{q_1}+ z \lmd_{q_1}\td\bt_{q_1},\nb\\
p_{\hat q_2}&=&\lmd_{q_2}\td\lmd_{q_2}+\bt_{q_2}\td\bt_{q_2}-z \lmd_{q_1}\td\bt_{q_1},
\eea
and
\be\label{OrthCon}
\langle\lmd_{q_1},\lmd_{q_2}\rangle[\td\bt_{q_1},\td\lmd_{q_2}]+ \langle\lmd_{q_1},\bt_{q_2}\rangle[\td\bt_{q_1},\td\bt_{q_2}]=0.
\ee
It is easy to check that the momentum conservation are met and the external field are still on-shell after the shifting. Hence, the corresponding amplitudes are the on-shell scattering amplitude of particles with complex momenta. They can be computed by the usual Feynman rules. If the amplitudes are constructible, they can also be computed by the BCFW recursion relations. For two massive fields, the spinors are linear functions of $z$. According to (\ref{twoLS}), the shift is accomplished by
\bea
\bt_{q_1}\rightarrow\hat\bt_{q_1}&=&\bt_{q_1}+z\lmd_{q_1},\nb\\
\lmd_{q_2}\rightarrow\hat\lmd_{q_2}&=&\lmd_{q_2}-z c_1 \lmd_{q_1},\nb\\
\bt_{q_2}\rightarrow\hat\bt_{q_2}&=&\bt_{q_2}-z c_2 \bt_{q_1},
\eea
and the other spinors in the two labeled particles are left invariant. Here $c_1={[\td\bt_{q_2},\td\bt_{q_1}]\over m},c_2=-{[\td\lmd_{q_2},\td\bt_{q_1}]\over m}$ and $\td\bt_{q_1}=c_1\td\lmd_{q_2}+c_2\td\bt_{q_2}.$

The spin states of these lines can be selected to be independent of $z$ after shifting. For the momentum shift (\ref{twoLS}), the $z$-independent states for two labeled particles are
\be\label{indeWave}
{\lmd_{q_1}\choose \td\bt_{q_1}},  {a\choose \td b}={1\over c_1} {\hat\lmd_{q_2}\choose \td\bt_{q_2}}-{1\over c_2} {\hat\bt_{q_2}\choose -\td\lmd_{q_2}}.
\ee

Hence when external particles are all massive dirac field, we can choose all the external states to be $z$-independent even after shifting.

Now we consider the case that one of the external line is  massless gauge field of $+$ helicity. Then the amplitudes are constructible under the two-line shift \cite{Badger1,Badger2}
\bea\label{pheliS}
p_1&=&\lmd_1\td\lmd_1+ z ([\td\lmd_2,\td\lmd_1]\lmd_2 \td\lmd_1+ [\td\bt_2,\td\lmd_1]\bt_2\td\lmd_1),\nb\\
p_2&=&\lmd_2\td\lmd_2+\bt_2\td\bt_2-z ([\td\lmd_2,\td\lmd_1]\lmd_2 \td\lmd_1+ [\td\bt_2,\td\lmd_1]\bt_2\td\lmd_1),
\eea
where $p_1$ momentum of the massless gauge boson and $p_2$ is for the dirac fields.
For the gauge field of positive helicity, the external field are of form
\be
\epsilon^+={\mu\td\lmd_1 \over\langle\mu,\lmd_1\rangle}.
\ee
Under this momentum shift, the gauge field transforms as
\be
\epsilon^+={\mu\td\lmd_1 \over  \langle\mu,\lmd_1\rangle+z [\td\lmd_2,\td\lmd_1] \langle\mu,\lmd_2\rangle+z[\td\bt_2,\td\lmd_1]\langle\mu,\bt_2 \rangle}.
\ee
The $z$-independent wave-function of the shift massive dirac field is
\be\label{indeWaveP}
{a\choose \td b}={1\over c_1} {\lmd_{q}\choose \hat{\td{\bt}}_{q}}+{1\over c_2} {\bt_{q}\choose -\hat{\td{\lmd}}_{q}},
\ee
where $c_1=[\td\bt_q,\td\lmd_g], c_2=[\td\lmd_q,\td\lmd_g]$
and $\hat{\td{\lmd}}_q=\td\lmd_q-z[\td\lmd_q,\td\lmd_g]\td\lmd_g, \hat{\tilde{\bt}}_q=\td\bt_q-z[\td\bt_q,\td\lmd_g]\td\lmd_g.
$

If the massless gauge field is of $-$ helicity with spinor form $\epsilon^-={\lmd_1\td\mu \over [\td\lmd_1 ,\td\mu]}$, then we can choose to shift the momenta as in \cite{Badger1,Badger2}
\bea\label{nheliS}
p_1&=&\lmd_1\td\lmd_1+ z (\langle\lmd_1,\lmd_2\rangle\lmd_1 \td\lmd_2+ \langle\lmd_1,\bt_2\rangle\lmd_1\td\bt_2),\nb\\
p_2&=&\lmd_2\td\lmd_2+\bt_2\td\bt_2-z (\langle\lmd_1,\lmd_2\rangle\lmd_1 \td\lmd_2+ \langle\lmd_1,\bt_2\rangle\lmd_1\td\bt_2),
\eea
Then the gauge field of negative helicity transforms as
\be\label{indeWavenG}
\epsilon^-={\lmd_1\td\mu \over [\td\lmd_1 ,\td\mu]+z \langle\lmd_1,\lmd_2\rangle[\td\lmd_2 ,\td\mu]+z \langle\lmd_1,\bt_2\rangle [\td\bt_2,\td\mu]}.
\ee
And the $z$-independent wave function is
\be\label{indeWaveP}
{a\choose \td b}={1\over c_1} {\hat\lmd_{q}\choose \td{\bt}_{q}}-{1\over c_2} {\hat\bt_{q}\choose -\td{\lmd}_{q}},
\ee
where $c_1=\langle\lmd_g,\lmd_q\rangle, c_2=\langle\lmd_g,\bt_{q}\rangle,$ and $\hat\lmd_q=\lmd_q-z\langle\lmd_g,\lmd_q\rangle\lmd_g, \hat\bt_q=\bt_q-z\langle\lmd_g,\bt_q\rangle\lmd_g.$
According to the discussion above, we find it is possible to choose proper momentum shift scheme and spin states such that the external states are z-independent and proportional  to $1\over z$ when $z\rightarrow\infty$ for the dirac fields and gluons respectively.

Under these conventions, we can argue that the amplitudes in QCD can are constructible when choosing proper spin configuration and momentum shift schemes. Let's denote the shifted amplitudes as $A(z)$. In general there are three cases. First, the external lines are of two or more massless gauge fields. Second, external lines contain only one massless gauge field. Third, external lines are all massive dirac fields.

For the first case, it is possible to choose to shift the momentum of  two massless gauge bosons. The amplitude are constructive and the recursion relations for the amplitudes are well-known  in \cite{Badger1,Badger2} for several years.

For the second case, we can select one of the shifted lines to be  the dirac field and the other to be gauge boson.  As discussed above, the massive fields and the massless can be choose to be z-independent and to tend as $1\over z$ respectively. Actually, as discussed in the documents \cite{Badger1}, the Feynman diagram with most high z-dependence are those in which all vertices along the complex path are trivalent composed of pure gauge fields. A complex path made up of such $r$ vertices has $r+1$ gauge boson propagators. Hence the amplitude at most tend to $z^r/z^{r+1}=1/z$ as $z\rightarrow \infty$. Furthermore, we can also choose two quark fields or two anti-quark fields to be the shifted line. The behavior of amplitude with respect to $z$ is the same as the third case as follows.

For the third case, since we shifted two quark fields or two anti-quark fields, the $z$-dependence flows pass through at least one internal gluon line. Furthermore, the external field can be chosen to be $z$-independent. Hence the amplitudes in QCD of this spin configuration  behave at most as ${1\over z}$ when $z\rightarrow \infty$.

We call the amplitudes which are tend to vanish  as directly constructive. The amplitudes with any spin structures will be studied in Section \ref{Sec-Gamp}. There are simple recursion relations for these constructive amplitudes \cite{Badger1,Badger2}
\be\label{twoShift}
\mathcal{A}(p_1,\cdots p_n)=\sum_{pt}\sum_h \mathcal{A}_L(p_r,\cdots \hat{p}_i,\cdots,p_s,-\hat{P}^h){1\over P_{ij}^2-m_{P_ij}^2}\mathcal{A}_R(\hat{P}^h, p_{s+1},\cdots \hat{p}_j,\cdots,p_{r-1}),
\ee
where summation is over all partitions of the external fields. $z$ can be solved from the on-shell condition of the intermediate momentum \cite{Badger1,Badger2}
\be\label{zvalue}
z={P^2-m_p^2\over -2P\cdot \eta},
\ee
where $\eta$ is the momentum of the $z$-term in the shifted momentum.

Actually, it is also possible to use three-line shift scheme.
We shift the two quark lines and the other lines can be either gauge field or dirac field. In the former case, the momentum shift as \cite{cohen}
\bea\label{Momshift}
p_1&=&\lmd_1\td\lmd_1+\bt_1\td\bt_1+a z \lmd_1 \td\bt_1,\nb\\
p_2&=&\lmd_2\td\lmd_2+\bt_2\td\bt_2+b z \lmd_1 (\langle\lmd_2,\lmd_1\rangle \td\lmd_2 +\langle\bt_2,\lmd_1\rangle\td\bt_2), \nb\\
p_3&=&\lmd_3\td\lmd_3+\bt_3\td\bt_3+c z \lmd_1 (\langle\lmd_3,\lmd_1\rangle \td\lmd_3 +\langle\bt_3,\lmd_1\rangle\td\bt_3),
\eea
where $a,b,c$ are constants such that
\be\label{Momcon}
a\td\bt_1+b (\langle\lmd_2,\lmd_1\rangle \td\lmd_2 +\langle\bt_2,\lmd_1\rangle\td\bt_2)+ c(\langle\lmd_3,\lmd_1\rangle \td\lmd_3 +\langle\bt_3,\lmd_1\rangle\td\bt_3)=0.
\ee
We can check that they satisfy the shifts conditions: external particle on-shell condition, momentum conservation and all multi-particle invariants shift linearly in $z$ \cite{cohen}. Since the three momentums are independent with each other for general momentums, the Eq. \ref{Momcon} can be satisfied by choosing suitable constants $a,b,c$.

It is easy to get the shift of the external massive dirac field. The spinor form of the dirac fields of momentum $p=\lmd\td\lmd+\bt\td\bt$ are
\be\label{wavefun}
u^{-}={\lmd\choose \td\bt},~~u^{+}={-\bt\choose \td\lmd}, ~~v^-={-\lmd\choose \td\bt}, ~~v^+={\bt\choose \td\lmd}.
\ee
After shifting, the spinor of particle 1 can be chosen to transforms as
\bea\label{spinshiftSim}
\td\lmd_1&\rightarrow& \td\lmd_1+ a z \td\bt_1,\nb\\
\td\bt_1&\rightarrow& \td\bt_1,\nb\\
\lmd_1&\rightarrow&\lmd_1,\nb\\
\bt_1&\rightarrow& \bt_1.
\eea

This can be seen directly to write the momentum into the matrix form as \bea
(\lmd_1, \bt_1)\left(
\begin{array}{cc}
 1 & a z \\
 0 & 1
\end{array}
\right)
\left(
\begin{array}{c}
 \td\lmd_1 \\
 \td\bt_1
\end{array}
\right).
\eea

Similarly, the spinors of the particle 2 and 3 transforms under the shift as
\bea\label{spinshiftGen}
\td\lmd_i&\rightarrow& \td\lmd_i+ {b z\over m}\langle\lmd_1,\bt_i\rangle \left(\langle\lmd_1,\lmd_i\rangle\td\lmd_i+\langle\lmd_1,\bt_i\rangle \td\bt_i\right),\nb\\
\td\bt_i&\rightarrow& \td\bt_i- {b z\over m}\langle\lmd_1,\lmd_i\rangle \left(\langle\lmd_1,\lmd_i\rangle\td\lmd_i+\langle\lmd_1,\bt_i\rangle \td\bt_i\right),\nb\\
\lmd_i&\rightarrow&\lmd_i,\nb\\
\bt_i&\rightarrow& \bt_i.
\eea

According to (\ref{wavefun}), (\ref{spinshiftSim}), (\ref{spinshiftGen}), we find it is possible to choose a particular linear combination for the two independent solutions $u_{\pm}$ or $v_{\pm}$ such that the wave functions are invariant under the shift. For example, we can choose $u^-$ for the particle 1.

For three-line shift, we should also sum of the three partitions of the shifted lines
\bea\label{threeShift}
\mathcal{A}(p_1,\cdots p_n)&=&\sum_{pt}\sum_h \mathcal{A}_L(p_r,\cdots \hat{p}_i,\hat{p}_j\cdots,p_s,-\hat{P}_{ij}^h){1\over P^2-m_P^2}\mathcal{A}_R(\hat{P}_{ij}^h, p_{s+1},\cdots \hat{p}_k,\cdots,p_{r-1})\nb\\
&+&\sum_{pt}\sum_h \mathcal{A}_L(p_r,\cdots \hat{p}_j,\hat{p}_k\cdots,p_s,-\hat{P}_{jk}^h){1\over P_{jk}^2-m_{P_{jk}}^2}\mathcal{A}_R(\hat{P}_{jk}^h, p_{s+1},\cdots \hat{p}_i,\cdots,p_{r-1})\nb\\
&+&\sum_{pt}\sum_h \mathcal{A}_L(p_r,\cdots \hat{p}_i,\hat{p}_k\cdots,p_s,-\hat{P}_{ik}^h){1\over P_{ik}^2-m_{P_{ik}}^2}\mathcal{A}_R(\hat{P}_{ik}^h, p_{s+1},\cdots \hat{p}_j,\cdots,p_{r-1}).\nb\\
\eea

We can also use a scheme with more shifted lines. Then the number of residues contribution to the amplitudes will  become larger as we shift more external lines. In this article, we only choose the most efficient way.

\section{The General procedure for the tree-level QCD amplitudes}\label{Sec-Gamp}
At tree level, the QCD amplitude are generally constructible  when choosing some particular spin configurations. However, for arbitrary spin configurations, the amplitudes are not usually directly constructible. In fact it depends on the shift schemes and also the varieties of  the external particles.
\begin{itemize}
\item
The amplitudes with two or more massless gluons are constructible when we choose two of the gluons as the shifted lines. The shift schemes depend on the helicity configuration of the two gluons. To get the amplitude, we can first fix the spin configuration for the massive dirac fields. Then we use the recursion relations to obtain the amplitude with this spin configuration. The amplitude with any spin configuration can be obtained by acting the little group generators on to each each external field.

\item
If there are only one massless gluon in the external lines of the amplitudes, we should at least choose one massive field as the shifted line and the other shifted line is the gluon. Then the amplitudes are constructible only when the spin state of the shifted line are z-independent after the momentum shifting. Such spin states are related to the gluon spinor. Worse than all, the spinors of the spin states depend on the momentum of the shifted particle. Hence its external states are invariant under the little group generators of the particle. That means we can not get the other spin states by acting with the generators of the little group.  As a result, we can not obtain the amplitude with another spin state for the shifted line by simply acting with the little group generators.

Instead, to get the amplitude with another spin state for this massive line $l_c$ (with other external states left unchanged), we can choose the two shifted lines as follows.  One of them is just this massive line. Another is a massive line $l_f$ which can not annihilate into a gluon with $l_c$. This time, the momentum shift bases on (\ref{twoLS}).  Since the amplitude is invariant under Lorentz group, we can choose a reference such that the spin state of the line $l_f$ in the former shift scheme are z-independent automatically. Then according to the analysis in Section \ref{MDfield}, the amplitude is still constructible. Hence we can get the amplitude with another spin state for $l_c$.

Linear combining the amplitudes obtained form two different shift schemes, we can get the amplitudes with all massive field of independent and little group sensitive spin states. Then we can act the little group generators on the amplitudes to get the amplitude with general configurations.

\item
Similar to the second case, with only massive external lines, we can also obtain the amplitudes with arbitrary helicity and spin configurations. But in the case, both of the shift schemes have to be chosen such that the shifted lines are two massive fields.
\end{itemize}

Our final aim is to get a general procedure for arbitrary tree-level amplitudes in QCD with massive dirac fields. To this end, we first list the three-parton amplitudes which are fully determined by the space-time symmetry and the gauge invariance for the gluons. Then we use the recursion relations to relate any tree level amplitudes to the three-parton amplitudes.  For convenience, we furthermore assume that the quarks are of different flavors in the amplitude. However, it is easy to extend out method for the amplitude include  the same flavor. In this article, the momentum of each particle is taken as incoming.
\subsection{Three parton amplitudes}\label{Sec-AmponeQoneG}
As explained in Section \ref{MDfield}, all the tree level amplitudes can be reduced to the three-parton amplitudes. We first show the formation of the three-parton amplitudes in QCD.
There are two different kinds of diagram for the amplitudes
\begin{figure}[htb]
\centering
\includegraphics{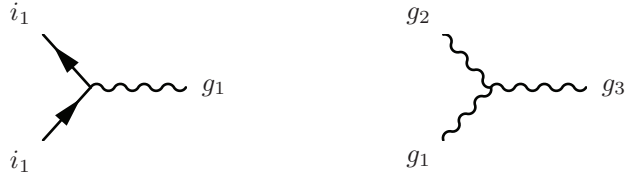}
\caption{Three-parton amplitudes in QCD.}
\end{figure}

Actually, the general spinor form of these amplitudes have been analyzed  in \cite{chen} for the massive case and in \cite{Cachazo1} for the massless case. For the amplitudes $A_{\bar q qg}$ corresponding to the former diagram, we denote the momentum as
\bea\label{Momshift}
p_{\bar q}&=&\lmd_{\bar q}\td\lmd_{\bar q}+\bt_{\bar q}\td\bt_{\bar q},\nb\\
p_q&=&\lmd_q\td\lmd_q+\bt_q\td\bt_q, \nb\\
p_g&=&\lmd_g\td\lmd_g,
\eea
For the three point amplitude $A_{ggg}$ with pure gluons, the momentum are taken as
\bea\label{Momshift}
p_{g_1}&=&\lmd_{g_1}\td\lmd_{g_1},\nb\\
p_{g_2}&=&\lmd_{g_2}\td\lmd_{g_2}, \nb\\
p_{g_3}&=&\lmd_{g_3}\td\lmd_{g_3},
\eea
Here we only need the tree level results. For the quark-gluon amplitudes, the formations are
\be\label{qqgAmp}
\bary{rclrcl}
A_{\bar q qg}({-1\over 2}, {-1\over 2}, -1)&=&{[\td\bt_{\bar q},\td\bt_q]\langle\lmd_g,\lmd_q\rangle \over [\td \bt_q, \td \lmd_g]}, ~~~&A_{\bar q qg}({-1\over 2}, {-1\over 2}, 1)&=&{\langle\lmd_{\bar q},\lmd_q\rangle[\td\bt_q,\td\lmd_g] \over \langle \lmd_q, \lmd_g\rangle}\nb\\
A_{\bar q qg}({1\over 2}, {-1\over 2}, -1)&=&{[\td\lmd_{\bar q},\td\bt_q]\langle\lmd_g,\lmd_q\rangle \over [\td \bt_q, \td \lmd_g]},~~~&A_{\bar q qg}({1\over 2}, {-1\over 2}, 1)&=&-{\langle\bt_{\bar q},\lmd_q\rangle[\td\bt_q,\td\lmd_g] \over \langle \lmd_q, \lmd_g\rangle}\nb\\
A_{\bar q qg}({-1\over 2}, {1\over 2}, -1)&=&{[\td\bt_{\bar q},\td\lmd_q]\langle\lmd_g,\lmd_q\rangle \over [\td \bt_q, \td \lmd_g]},~~~&A_{\bar q qg}({-1\over 2}, {1\over 2}, 1)&=&-{\langle\lmd_{\bar q},\bt_q\rangle[\td\bt_q,\td\lmd_g] \over \langle \lmd_q, \lmd_g\rangle}\nb\\
A_{\bar q qg}({1\over 2}, {1\over 2}, -1)&=&{[\td\lmd_{\bar q},\td\lmd_q]\langle\lmd_g,\lmd_q\rangle \over [\td \bt_q, \td \lmd_g]},~~~&A_{\bar q qg}({1\over 2}, {1\over 2}, 1)&=&{\langle\bt_{\bar q},\bt_q\rangle[\td\bt_q,\td\lmd_g] \over \langle \lmd_q, \lmd_g\rangle}\nb\\
\eary
\ee

The three-gluon amplitudes are given by the standard $MHV$ and $\bar{MHV}$ expressions \cite{Cachazo:2004kj,Cachazo1,Badger2}:
\be\label{gggAmp}
A_{ggg}(-,-,+)={\langle 12\rangle^3 \over \langle 32 \rangle \langle 31 \rangle}, ~~~~A_{ggg}(+,+,-)=-{[ 12]^3 \over [32] [31]}
\ee

\subsection{Amplitudes with two quark-antiquark pairs }\label{Sec-AmptwoQ}
As a warming up exercise, we apply the general procedure to the amplitudes with two massive quark-antiquark pairs. As discussed above, to get the amplitude for the massive spin particles with arbitrary spin configuration, we need choose two different kinds of shifted schemes. Here, one is to choose two  external quark fields as the shifted particles. The other is to choose one quark field and one anit-quark field with different flavor as the shifted fields.

First, we label the two quarks as the shifted particles $\hat{q}_1,\hat{q}_2$. Then the momentum shift as (\ref{twoLS}). We exhibit that the amplitudes with one particular spin configuration can be constructible. From the discussion in Section \ref{MDfield}, to make the amplitude constructible, the spin configuration for the two shifted particle should be chosen as ${\lmd_{q_1}\choose \td\bt_{q_1}},  {a\choose \td b}= {-p_{q_2}\circ\td\bt_{q_1}\choose m\td\bt_{q_1}}$.  Here we use the symbol $\circ$ to denote the inner product for the spinor index. For example, $p_{q_2}\circ\td\bt_{q_1}=\lmd_{q_2}[ \td\lmd_{q_2},\td\bt_{q_1}]+\bt_{q_2}[\td\bt_{q_2},\td\bt_{q_1}]$.
For the other two external fields, the spins are arbitrary. For the moment, we choose the spin to be ${-1\over 2}$ with respect to the momentum for each un-shifted field.

We denote the gauge invariant sub-amplitude with this particular spin configuration as $\mathcal{A}({\hat q_1}^{-1\over 2},{\bar q_1}^{-1\over 2},{\hat q_2}^{z},{\bar q_2}^{-1\over 2})$. The external waves of the fields are ${\lmd_{q_1}\choose \td\bt_{q_1}}$, $(\lmd_{\bar q_1},\td\bt_{\bar q_1})$, ${a\choose \td b}$, $(\lmd_{\bar q_2},\td\bt_{\bar q_2})$ respectively. The diagram contribute to the amplitude is

\begin{figure}[htb]
\centering
\includegraphics{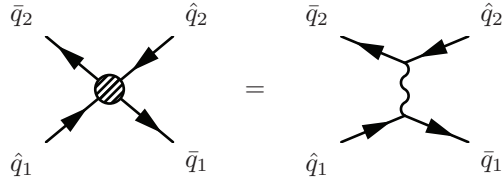}
\caption{Amplitudes with two pairs of dirac fields.}
\end{figure}

According to the on-shell recursion relation, we can get
\begin{eqnarray}\label{twolineRec}
&&\mathcal{A}({\hat q_1}^{-1\over 2},{\bar q_1}^{-1\over 2},{\hat q_2}^{z},{\bar q_2}^{-1\over 2})= \sum_h \mathcal{A}_L(\bar q_1, \hat q_1,g_1^{-P_{12},-h}){1\over P^2}\mathcal{A}_R(g_1^{P_{12},h}, \bar q_2,\hat q_2) \nb\\
&=&{2\over (p_{q_1}+p_{\bar q_1})^2}\left({[\td\bt_{\bar q_1},\td\bt_{q_1}]\langle\lmd_{q_1},\hat\lmd_g\rangle \over [\hat{\td \lmd}_g,\td \bt_{q_1}]} {\langle\lmd_{\bar q_2},a\rangle[\td b,\hat{\td\lmd}_g] \over \langle a, \hat\lmd_g\rangle}\right)
+{2\over (p_{q_1}+p_{\bar q_1})^2}\left({\langle\lmd_{\bar q_1},\lmd_{q_1}\rangle[\td\bt_{q_1},\hat{\td\lmd}_g] \over \langle \hat\lmd_g,\lmd_{q_1}\rangle} {[\td\bt_{\bar q_2},\td b]\langle a,\hat\lmd_g \rangle \over [\td b, \hat{\td\lmd}_g]}\right)\nb\\
&=&{2\over (p_{q_1}+p_{\bar q_1})^2}\left({[\td\bt_{\bar q_1},\td b]\langle\hat\lmd_g,\lmd_{q_1}\rangle \langle\lmd_{\bar q_2},a\rangle\over \langle a, \hat\lmd_g\rangle}\right)
+{2\over (p_{q_1}+p_{\bar q_1})^2}\left({\langle\lmd_{\bar q_1},\lmd_{q_1}\rangle [\td\bt_{\bar q_2},\td \bt_{q_1}]\langle\hat\lmd_g,a \rangle\over \langle \lmd_{q_1}, \hat\lmd_g\rangle} \right).
\end{eqnarray}
Such amplitude can also be obtained from the Feynman rules
\bea\label{fourPointFeyn}
\mathcal{A}({\hat q_1}^{-1\over 2},{\bar q_1}^{-1\over 2},{\hat q_2}^{z},{\bar q_2}^{-1\over 2})&=&{2\over (p_{q_1}+p_{\bar q_1})^2}\left(\langle\lmd_{\bar q_1}, \lmd_{\bar q_2}\rangle[\td\bt_{q_1},\td b]+\langle\lmd_{\bar q_1}, a\rangle [\td\bt_{q_1},\td\bt_{\bar q_2}]\right.\nb\\
&&\left.+\langle\lmd_{\bar q_2},\lmd_{q_1}\rangle [\td b,\td\bt_{\bar q_1}]+\langle a, \lmd_{q_1}\rangle [\td\bt_{\bar q_2},\td\bt_{\bar q_1}]\right)\nb\\
&=&{2\over (p_{q_1}+p_{\bar q_1})^2}\left(\langle\lmd_{\bar q_1}, a\rangle [\td\bt_{q_1},\td\bt_{\bar q_2}]+\langle\lmd_{\bar q_2},\lmd_{q_1}\rangle [\td b,\td\bt_{\bar q_1}]\right).
\eea
The second equality is deduced according to the equation (\ref{OrthCon}) and (\ref{indeWave}) since here we choose a particular reference frame and the external labeled particles are $z$-independent even after shift. The result obtained from two method are consistent. To see this, we can insert the identity operator $\mathbb{I}_L={,\lmd_2\rangle \, \langle\lmd_1,-,\lmd_1\rangle \, \langle\lmd_2,\over \langle\lmd_1,\lmd_2\rangle}$ into the  inner product of the left spinors in (\ref{fourPointFeyn}). Then the inner product can be written as
\bea\label{InsertID}
\langle\lmd_{\bar q_2}, \lmd_{q_1}\rangle&=&{-\langle\lmd_{\bar q_2}, a\rangle\langle\lmd_{g}, \lmd_{q_1}\rangle +\langle\lmd_{\bar q_2}, \lmd_{g}\rangle\langle a, \lmd_{q_1}\rangle \over \langle a, \lmd_{g}\rangle} , \nb\\
\langle\lmd_{\bar q_1}, a\rangle&=& {\langle\lmd_{\bar q_1}, \lmd_{g}\rangle\langle\lmd_{q_1}, a\rangle- \langle\lmd_{\bar q_1}, \lmd_{q_1}\rangle\langle\lmd_{g}, a\rangle \over \langle\lmd_{q_1}, \lmd_{g}\rangle}.
\eea
Combining the equations (\ref{InsertID}) and (\ref{fourPointFeyn}), we can find the amplitude obtained from the Feynman rules are identical with the results by on-shell recursion relations for the given spinor structure.

Since the spin state of the 1th-quark field depend on the spinor of the 2th-quark field, we do not get the amplitude for general spin configurations. As the discussion above, we can not acting on it with little group generators directly to get the amplitude for any spin configuration.

Before that, we should consider the second momentum shift scheme. We label 1st-antiquark field and 2nd-quark field as shifted particles. The momentums of them are shifted as (\ref{twoLS}). And the external wave-functions for the shifted fields are taken as $(\lmd_{\bar q_1}, \td\bt_{\bar q_1}),  {a\choose \td b}= {-p_{q_2}\circ\td\bt_{\bar q_1}\choose m\td\bt_{\bar q_1}}$. While the wave-fucntions of the other two un-shifted particles are still taken as ${\lmd_{q_1}\choose \td\bt_{q_1}}$, $(\lmd_{\bar q_2},\td\bt_{\bar q_2})$. Such kind of amplitude is still constructible.

The constructible amplitude, in which  all external field are z-independent after momentum shift, is
\bea\label{Ampqqqq2}
\mathcal{A}({q_1}^{-1\over 2},{\hat{\bar q}_1}^{-1\over 2},{\hat q_2}^{z},{\bar q_2}^{-1\over 2})
&=&\sum_h \mathcal{A}_L(\hat{\bar q}_1,  q_1,g_{-P_{12},-h}){1\over P_{12}^2}\mathcal{A}_R(g_{P_{12},h}, \bar q_2,\hat q_2)\nb\\
&=&{2\over (p_{q_1}+p_{\bar q_1})^2}\left(\langle\lmd_{ q_1}, a\rangle [\td\bt_{\bar q_1},\td\bt_{\bar q_2}]+\langle\lmd_{\bar q_2},\lmd_{\bar q_1}\rangle [\td b,\td\bt_{q_1}]\right).
\eea

For general external momentums, the z-independent spin states for the 2nd-quark field in above two shift scheme are different. Furthermore, the spin states of other field are invariant. Since the amplitude are linear with respect to the external waves. We can recover the amplitude with general spin configurations for the 2nd-quark field by the linear combination of the two constructible amplitude.  And for the independent spin state $(-{1\over 2})$ of the 2nd-quark field, we can also obtain the amplitude
\bea
&&\mathcal{A}({q_1}^{-1\over 2},{\bar q_1}^{-1\over 2},{ q_2}^{-1\over 2},{\bar q_2}^{-1\over 2})\nb\\
&=&{2\over (p_{q_1}+p_{\bar q_1})^2}\left(\langle\lmd_{\bar q_1}, \lmd_{\bar q_2}\rangle[\td\bt_{q_1},\td \bt_{q_2}]+\langle\lmd_{\bar q_1}, \lmd_{q_2}\rangle [\td\bt_{q_1},\td\bt_{\bar q_2}]\right.
\left.+\langle\lmd_{\bar q_2},\lmd_{q_1}\rangle [\td \bt_{q_2},\td\bt_{\bar q_1}]+\langle \lmd_{q_2}, \lmd_{q_1}\rangle [\td\bt_{\bar q_2},\td\bt_{\bar q_1}]\right)\nb\\
\eea
Then the amplitude with any spin configuration can be obtained by the actions of little group generators for each external line.
\bea\label{AmpGenSpin}
&&\mathcal{A}({\pm 1\over 2}_{q_1},{\pm 1\over 2}_{\bar q_2},{\pm 1\over 2}_{q_2},{\pm 1\over 2}_{\bar q_1})\nb\\
&=&(J^+_{q_1})^{n_{q_1}}(J^+_{\bar q_2})^{n_{\bar q_2}} (J^+_{q_2})^{n_{q_2}} (J^+_{\bar q_1})^{n_{\bar q_1}}{\mathcal{A}({- 1\over 2}_{q_1},{-1\over 2}_{\bar q_2},{-1\over 2}_{q_2},{- 1\over 2}_{\bar q_1})\over (p_{q_1}+p_{\bar q_1})^2},
\eea
where $n=0,1$ when the spin equals $\pm{1\over 2}$ respectively. The amplitude obtained here are consistent with that get by usual Feynman rules.

\subsection{Amplitudes with one quark pair and two gluons}\label{Sec-AmponeQtwoG}
These kinds of amplitudes were considered several times before \cite{Badger2,Ozeren}. Here we compute it in a different momentum shift scheme to verify the validity of our method.

We first consider the amplitude with helicity configuration $(1^-, 2^-)$ for two external gluons and spin configuration $\hat q^{z},\bar q^{-1\over 2}$ for external massive lines. The hat is denoted for the shift line. As discussed in section \ref{MDfield}, we shift the external momentum as (\ref{nheliS}) in this case. Here the wave functions  are $\epsilon_1^- = {\lmd_1 \td u_1\over [\hat{\td\lmd}_1,\td u_1]},\epsilon_2^- = {\lmd_2 \td u_2\over [\hat{\td\lmd}_2,\td u_2]}$ for the  incoming two gluons and  ${a\choose \td b},{\lmd_{\bar q}\choose \td\bt_{\bar q}}$ for the two massive dirac lines, where ${a\choose \td b}$, which is defined in (\ref{indeWaveP}), is z-independent state under the momentum shift . The Feynman diagrams for this process is shown in Fig.(\ref{FigoneQtwoG}).

\begin{figure}[htb]
\centering
\includegraphics{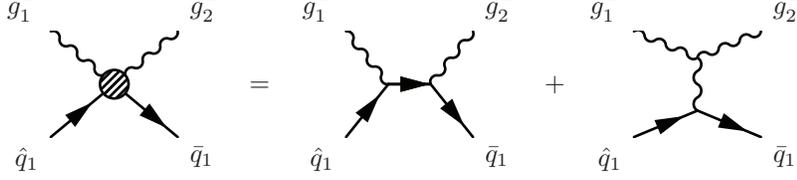}
\caption{Decomposition of the four-parton amplitude.}
\label{FigoneQtwoG}
\end{figure}

As the analysis in section \ref{MDfield}, the amplitude with this helicity and spin configuration are constructible if we choose the momentum shift as (\ref{nheliS}). Then according to the BCFW relation, the amplitude can be determined by  the simple poles and the corresponding residues. Furthermore, The former diagram in the right hand does not contribute to the residues. Hence only the second diagram contribute a term to the amplitude. According to the recursion relation, the amplitude are simply
\begin{eqnarray}\label{twolineRec}
\mathcal{A}(\hat q_1,\hat g^-_1, g^-_2,\bar{q}_1) &=& \sum_h \mathcal{A}_L(\bar q_1,\hat q_1, \hat g^{-\hat P_{q_1\bar q_1},-h}){1\over P_{q_1\bar q_1}^2}\mathcal{A}_R(\hat g^{\hat P_{q_1\bar q_1},h}, \hat g^-_1,g^-_2)\nb\\
&=&\mathcal{A}_L(\bar q_1,\hat q_1, \hat g^{-\hat P_{q_1\bar q_1},-}){1\over P_{q_1\bar q_1}^2}\mathcal{A}_R(\hat g^{\hat P_{q_1\bar q_1},+}, \hat g^-_1,g^-_2)\nb\\
&=&{1\over P^2_{q_1\bar q_1}}\left({[\td\bt_{\bar q},\td b]\langle \hat\lmd_{g_P}, a\rangle\over [\td b, \hat{\td \lmd}_{g_P}]} {\langle \lmd_{g_1},\lmd_{g_2}\rangle^3 \over \langle \hat\lmd_{g_P},\lmd_{g_2}\rangle\langle \hat\lmd_{g_P},\lmd_{g_1}\rangle}\right) \nb\\
&=&{-1\over P^2_{q_1\bar q_1}}{[\td\bt_{\bar q},\td b]\langle a, \lmd_{g_2}\rangle\langle \lmd_{g_1}, \lmd_{g_2}\rangle\over [\td b, \td \lmd_{g_1}]}  \nb\\
\end{eqnarray}

We can also get the amplitude from the Feynman rule but with more calculation. The second diagram on right hand of Fig.\ref{FigoneQtwoG} contribute a term
\be
\bar v(\bar q_1)\gamma^{\mu} u(q_1) {g_{\mu\nu}\over (p_q+p_{\bar q})^2}\left[g^{\nu\rho}(p-p_{g_2})^{\sigma}+g^{\rho\sigma} (p_{g_2}-p_{g_1})^\sigma+g^{\sigma\nu}(p_{g_1}-p)^\rho\right] \varepsilon^-_\sigma(g_1)\varepsilon^-_\rho(g_2).
\ee
Here we can choose the arbitrary spinors $\td \xi_1, \td \xi_2$ in the two gluons to be $\td\lmd_2,\td\lmd_1$ respectively. Then this term can be simplified into
\be\label{Ampggqq2}
{\langle\lmd_{\bar q_1},\lmd_2\rangle[\td b,\td\lmd_2]+\langle a, \lmd_2\rangle[\td\bt_{\bar q_1},\td\lmd_2]\over [\td\lmd_1,\td\lmd_2]^2}.
\ee

The contribution of the first diagram on the right hand of Fig. (\ref{FigoneQtwoG}) is
\be
\bar v(\bar q_1) \gamma^\mu \varepsilon^-_\mu(g_2) {p\!\!\!/_q+p\!\!\!/_{g_1}+m \over (p_q+p_{g_1})^2-m^2}\gamma^\nu\varepsilon^-_\nu(g_1)u(q_1).
\ee

It can also be written into the spinor form
\be\label{Ampggqq1}
{\langle\lmd_{\bar q_1},\lmd_2\rangle[\td b,\td\lmd_2]\over [\td\lmd_2,\td\lmd_1][\td\lmd_1,\td\lmd_2]}+m {[\td\lmd_1,\td\bt_{\bar q_1}]\langle\lmd_2,\lmd_1\rangle [\td b,\td\lmd_2]\over [\td\lmd_2,\td\lmd_1][\td\lmd_1,\td\lmd_2]2p_q \cdot p_{g_1}}.
\ee
Adding (\ref{Ampggqq2}) and (\ref{Ampggqq1}), we obtain the amplitude
\bea\label{Ampggqqt}
A_{Feyn}&=&{\langle a, \lmd_2\rangle[\td\bt_{\bar q_1},\td\lmd_2]\over [\td\lmd_1,\td\lmd_2]^2}+m {[\td\lmd_1,\td\bt_{\bar q_1}]\langle\lmd_2,\lmd_1\rangle [\td b,\td\lmd_2]\over [\td\lmd_2,\td\lmd_1][\td\lmd_1,\td\lmd_2]2p_q \cdot p_{g_1}}\nb\\
&=&{\langle a, \lmd_2\rangle[\td\bt_{\bar q_1},\td\lmd_2]\over [\td\lmd_1,\td\lmd_2]^2}+m {\langle\lmd_2,\lmd_1\rangle [\td b,\td\lmd_2]\over [\td\lmd_2,\td\lmd_1]2p_q \cdot p_{g_1}}\left([\td b,\td\bt_{\bar q}]+{[\td b,\td \lmd_1][\td\lmd_2,\td\bt_{\bar q}]\over [\td\lmd_1,\td\lmd_2]}\right)\nb\\
&=&{\langle\lmd_2,a\rangle [\td\bt_{\bar q},\td b]\over [\td\lmd_1,\td\lmd_2]\rangle[\td b,\td\lmd_1]}.
\eea
It is easy to see the result are consistent with tho one from the BCFW reduction relation.

To get the amplitude with other spin configuration for the massive dirac lines, we also can not act with generators of the little group directly. Here, the spin state of the quark line are not independent but related with the spinors of a gluon. In fact the spin state for the quark field is ${a\choose \td b}={-m\lmd_1\choose \widetilde{\lmd_1\circ p_q}}$.

We should first get the amplitude with all  the massive external fields of independent spin states. To this end, we choose the 2-gluon and the quark field as the shifted line. This will help to get the amplitude with all external configuration invariant except for the quark field. The amplitude is constructive under this kind of shift only when the spin states of the quark field are related to the 2-gluon. Actually, the spin state of constructive amplitude have to be chosen as ${-m\lmd_2 \choose \widetilde{\lmd_2\circ p_q}}$. Then by the BCFW recursion relation, we can get the amplitude with the quark spin state changed while other spin or helicity configurations invariant.

If we shift the momentum of the quark field and 1-gluon field as (\ref{nheliS}), the amplitude, in which  quark spin state are ${a\choose \td b}={-m\lmd_2 \choose \widetilde{\lmd_2\circ p_q}}$, are constructible and hence can be obtained by the BCFW recursion relations. We denote this amplitude as $\mathcal{A}(\hat q^{z},1^-,\hat 2^-,\bar q_{-1\over 2})$. In this kind of shift scheme, both Feynman diagram contribute poles to the complexified amplitude. The second diagram in right hand of Fig.\ref{FigoneQtwoG} lead
\bea\label{BCFWggqq11}
\mathcal{A}_1&=&{1\over P^2_{q\bar q}}\left({[\td \bt_{\bar q}, \td b]\langle\hat\lmd_{g_P}, a\rangle \over [\td b,\hat{\td \lmd}_{g_P}]} {\langle \lmd_1, \lmd_2\rangle^3 \over \langle\hat\lmd_{g_P},\lmd_2\rangle\langle\hat\lmd_{g_P},\lmd_1 \rangle}\right)\nb\\
&=&{[\td\bt_q,\td b]\langle\lmd_1,a\rangle\over [\td b,\td\lmd_2][\td\lmd_1,\td\lmd_2]}.
\eea
And the first diagram leads
\bea\label{BCFWggqq12}
\mathcal{A}_2&=&{1\over P^2_{q g_1}}\left({\langle\lmd_{\bar q},\lmd_2\rangle[\td\bt_{q_P},\td\xi_2] +[\td\xi_2,\td\bt_{\bar q}]\langle\lmd_{2},\lmd_{q_P}\rangle\over [\hat{\td\lmd}_2,\td\xi_2]} \times {\langle\bt_{q_P},\lmd_1\rangle[\td b,\td\xi_1] +[\td\xi_1,\td\lmd_{q_P}]\langle\lmd_{1},a\rangle\over [\hat{\td\lmd}_1,\td\xi_1]}\right.\nb\\
&&+ \left. {\langle\lmd_{\bar q},\lmd_2\rangle[\td\lmd_{q_P},\td\xi_2] +[\td\xi_2,\td\bt_{\bar q}]\langle\lmd_{2},-\bt_{q_P}\rangle\over [\hat{\td\lmd}_2,\td\xi_2]} \times {\langle\lmd_{q_P},\lmd_1\rangle[\td b,\td\xi_1] +[\td\xi_1,-\td\bt_{q_P}]\langle\lmd_{1},a\rangle\over [\hat{\td\lmd}_1,\td\xi_1]}\right)\nb\\
&=&{1\over P_{q g_1}^2-m^2}{[\td b,\td\bt_{\bar q}] \langle\lmd_1,\lmd_2\rangle\langle\lmd_1,a\rangle\over[\td\lmd_2,\td b]}.
\eea
To deduce the second equality, we choose the most convenient gauge $\td\xi_1=\td\xi_2=\td b.$ From the Feynman rule, we can also get the amplitude for this spin and helicity configuration. The results is exactly the same as the one from the BCFW recursion relation
\bea\label{Ampggqq}
\mathcal{A}(\hat q^{z},1^-,\hat 2^-,\bar q_{-1\over 2})&=& {1\over P_{q g_1}^2-m^2}{[\td b,\td\bt_{\bar q}] \langle\lmd_1,\lmd_2\rangle\langle\lmd_1,a\rangle\over[\td\lmd_2,\td b]}+{1\over P_{g_2 g_1}^2}{[\td\bt_q,\td b]\langle\lmd_1,a\rangle\langle\lmd_1,\lmd_2\rangle\over [\td b,\td\lmd_2]}.
\eea

The amplitude discussed above are all of correlated spin states between the quark field with a gluon field. However, if the two gluons are not collinear, the spin state of the quark field are different for the two kinds of shift scheme. Hence, for any independent spin state for quark field, the amplitude can be obtained as a linear combination of the amplitudes which are constructive under the corresponding shift scheme. Then the amplitude with other spin configurations are directly to obtain by acting with the little group generators for each external massive lines. For example, the amplitude with spin and helicity configuration $(q^{-1\over 2},1^-, 2^-,\bar q^{-1\over 2})$ is
\bea\label{Ampqnhgggnh}
\mathcal{A}(q^{-1\over 2},1^-, 2^-,\bar q^{-1\over 2})&=& {\langle\lmd_2,\lmd_q\rangle\over m\langle\lmd_1,\lmd_2\rangle} \mathcal{A}(\hat q^{z},\hat 1^-, 2^-,\bar q_{-1\over 2}) -{\langle\lmd_1,\lmd_q\rangle\over m\langle\lmd_1,\lmd_2\rangle}
\mathcal{A}(\hat q^{z},1^-,\hat 2^-,\bar q_{-1\over 2}) \nb\\
&=&
{[\td\bt_{\bar q},\widetilde{\lmd_2\circ p_q}]\langle\lmd_1,\lmd_2\rangle\langle\lmd_1,\lmd_q\rangle +[\td\bt_{\bar q},\widetilde{\lmd_1\circ p_q}]\langle\lmd_2,\lmd_1\rangle\langle\lmd_2,\lmd_q\rangle\over 4 p_q\cdot p_{g_1} p_{g_2}\cdot p_{g_1}}.
\eea

The result obtained by our method is the same as the one in Feynman rules. For the amplitudes with other spin configurations, we can calculate them by acting with the little group generators for the quark field and the anti-quark field. For other helicity configuration for the gluons, we can obtain the amplitude similarly.
\bea\label{Ampqnhgggnh}
\mathcal{A}(q^{-1\over 2},1^-, 2^+,\bar q^{-1\over 2})
&=&
{\langle\lmd_1,p_q\circ\td\lmd_2\rangle\left(-{1\over m}\langle p_q\circ\td\lmd_2,\lmd_q\rangle\langle\lmd_{\bar q},\lmd_1\rangle+\langle\lmd_1,\lmd_q\rangle[\td\bt_{\bar q},\td\lmd_2]\right)\over 4 p_q\cdot p_{g_1} p_{g_2}\cdot p_{g_1}}\nb\\
\mathcal{A}(q^{-1\over 2},1^+, 2^-,\bar q^{-1\over 2})
&=&
{[\td\lmd_1,\widetilde{p_q\circ\lmd_2}]\left(-{1\over m}[ \widetilde{p_q\circ\lmd_2},\td\bt_q][\td\bt_{\bar q},\td\lmd_1]+[\td\lmd_1,\td\bt_q]\langle\lmd_{\bar q},\lmd_2\rangle\right)\over 4 p_q\cdot p_{g_1} p_{g_2}\cdot p_{g_1}}\nb\\
\mathcal{A}(q^{-1\over 2},1^+, 2^+,\bar q^{-1\over 2})
&=&
{\langle\lmd_{\bar q},\td\lmd_2\circ p_q\rangle[\td\lmd_1,\td\lmd_2][\td\lmd_1,\td\bt_q] +\langle\lmd_{\bar q},\td\lmd_1\circ p_q\rangle[\td\lmd_2,\td\lmd_1][\td\lmd_2,\td\bt_q]\over 4 p_q\cdot p_{g_1} p_{g_2}\cdot p_{g_1}}.
\eea
Similarly as (\ref{AmpGenSpin}), the amplitudes with other spin configurations for the quark and anti-quark lines can be obtained by acting with little group generator $\mathcal{R}(J^+)$  on them.
\subsection{Amplitudes with two quark pairs and one gluon}\label{Sec-AmptwoQoneG}
In this section, we consider the amplitudes with two quark pairs and one gluon for all the spin and helicity configurations. We have to choose one massive field as the shifted line $\hat q_1$. And we can select another field to be either the massless gluon $\hat g$ or another massive fields $\hat q_2$. As discussed in the second item at the beginning of this section, to obtain the full amplitude,   we need two different pairs of the shifted lines. One is of the shifted lines $\hat q_1$ and $\hat g$, the other is to shift the lines $\hat q_1$ and $\hat g$.

We first analysis the case when the shifted lines are $\hat q_1$ and $\hat g$. For the moment, we set the helicity of the gluon to be $+$ and the spin of other unshifted line are ${-1\over 2}$. We denote the sub-amplitude as $\mathcal{A}_R(\hat g^{-}, {\bar g_2}^{-1\over 2}, g_2^{-1\over 2}, {\bar g_1}^{-1\over 2}, \hat g_1^z)$. The momentum shift is accomplished according to (\ref{nheliS}). Hence the wave function of the shifted lines should be (\ref{indeWavenG}) and (\ref{indeWaveP}) for $\hat q_1$ and $\hat g$ respectively such that the amplitude can be constructible. Then external wave function are ${\lmd_g\td\xi\over [\hat{\td\lmd}_g,\td\xi]}$, $(\lmd_{\bar q_2},\td\bt_{\bar q_2})$,${\lmd_{q_2}\choose \td\bt_{q_2}}$, $(\lmd_{\bar q_1},\td\bt_{\bar q_1})$, ${a\choose \td b}$ respectively, where $\hat{\td\lmd}_g=\td\lmd_g+z \widetilde{\lmd_g\circ p_{q_1}}$ and ${a\choose \td b}={-m\lmd_g\choose \widetilde{\lmd_g\circ p_{q_1}}}$. In this case, the diagrams contribute to the amplitude are

\begin{figure}[htb]
\centering
\includegraphics{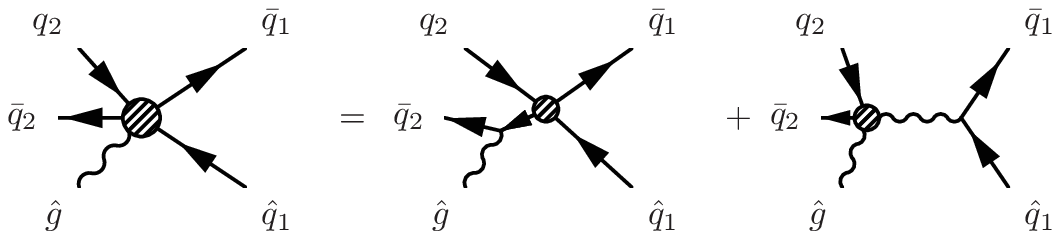}
\caption{Decomposition of the five-point amplitudes after shifting the lines $\hat g$ and $\hat q_1$.}
\end{figure}

According to the diagram, the amplitude can be expressed as
\begin{eqnarray}\label{twolineRec}
\mathcal{A}_R(\hat g^{-}, {\bar g_2}^{-1\over 2}, g_2^{-1\over 2}, {\bar g_1}^{-1\over 2}, \hat g_1^z) &=& \sum_s \mathcal{A}_L(\hat g^-, \bar q_2,\hat q_2^{-P_{g\bar q_2},-s}){1\over \hat P_{g\bar q_2}^2-m_2^2}\mathcal{A}_R(\hat q_2^{P_{g\bar q_2},s}, \bar q_2,\bar q_1, \hat q_1)\nb \\
&+&\sum_h \mathcal{A}_L(\hat g^-, \bar q_2, q_2, \hat g_{-P_{gq_2,-h}} ){1\over \hat P_{gq_2}^2}\mathcal{A}_R(\hat g_{P_{gq_2,h}},\bar q_1, \hat q_1).
\end{eqnarray}
According to the result in subsection \ref{Sec-AmponeQoneG},\ref{Sec-AmptwoQ} and \ref{Sec-AmponeQtwoG}, we can get the spinor form of the amplitude
\begin{eqnarray}\label{AmptwoQz1}
&&\mathcal{A}_R(\hat g^{-}, {\bar g_2}^{-1\over 2}, g_2^{-1\over 2}, {\bar g_1}^{-1\over 2}, \hat g_1^z) \nb\\
&=& {1\over P^2_{\hat g \bar q_2}-m_2^2} {2\over (\hat P_{\bar q_1 \hat q_1})^2} {\langle\lmd_{\bar q_2},\lmd_1\rangle\over [\hat{\td\lmd}_1,\td\bt_{\bar q_2}]}\nb\\
&\times&\left(\langle \lmd_{\bar q_1}|\hat P|\td\bt_{\bar q_2}][\td b_{q_1},\td\bt_{q_2}]+m\langle\lmd_{\bar q_1},\lmd_{q_2}\rangle[\td b_{q_1},\td\bt_{\bar q_2}]-[\td\bt_{\bar q_2}|\hat P|a_{q_1}\rangle[\td\bt_{q_2},\td\bt_{\bar q_1}]-m\langle\lmd_{q_2},a_{q_1}\rangle[\td\bt_{\bar q_1},\td\bt_{\bar q_2}]\right)\nb \\
&+& {1\over  P_{q_1\bar q_1}^2}{[\td\bt_{\bar q_2},\widetilde{\lmd_2\circ p_{q_2}}]\langle\lmd_I,\lmd_2\rangle\langle\lmd_I,\lmd_{q_2}\rangle +[\td\bt_{\bar q_2},\widetilde{\lmd_I\circ p_{q_2}}]\langle\lmd_2,\lmd_I\rangle\langle\lmd_2,\lmd_{q_2}\rangle\over 4 p_{q_2}\cdot \hat p_{g_I}  p_{g_2}\cdot \hat p_{g_I}}{\langle\lmd_{\bar q_1}, \lmd_{q_1}\rangle[\td\bt_{q_1},\td\lmd_I]\over \langle\hat\lmd_{q_1},\lmd_I\rangle}\nb\\
&+&{1\over  P_{q_1\bar q_1}^2} {[\td\lmd_I,\widetilde{p_{q_2}\circ\lmd_2}]\left(-{1\over m}[ \widetilde{p_{q_2}\circ\lmd_2},\td\bt_{q_2}][\td\bt_{\bar q_2},\td\lmd_I]+[\td\lmd_I,\td\bt_{q_2}]\langle\lmd_{\bar q_2},\lmd_2\rangle\right)\over 4 p_{q_2}\cdot \hat p_{g_I} p_{g_2}\cdot \hat p_{g_I}} {[\td\bt_{\bar q_1},\td\bt_{q_1}]\langle\lmd_I, \lmd_{q_1}\rangle\over [\td\bt_{q_1},\td\lmd_I]}.
\end{eqnarray}
Here and the following, the $z$ values are dependent the intermediate massive lines and the chosen momentum shift scheme according to (\ref{zvalue}).

The amplitude with other spin state can be obtained by shift the line $\hat q_2$ and $\hat q_1$. The momentum shift is accomplished by (\ref{twoLS}) and (\ref{OrthCon}). Then the z-independent spin states of them are $
{\lmd_{q_2}\choose \td\bt_{q_2}},  {a\choose \td b}= {-p_{q_1}\circ\td\bt_{q_2}\choose m\td\bt_{q_2}}$. Other external line are chosen as above. We denote this amplitude as $\mathcal{A}_R( g^{-}, {\bar g_2}^{-1\over 2}, \hat g_2^{-1\over 2}, {\bar g_1}^{-1\over 2}, \hat g_1^z)$. The diagram contribute
to the pole of the amplitude are

\begin{figure}[htb]
\centering
\includegraphics{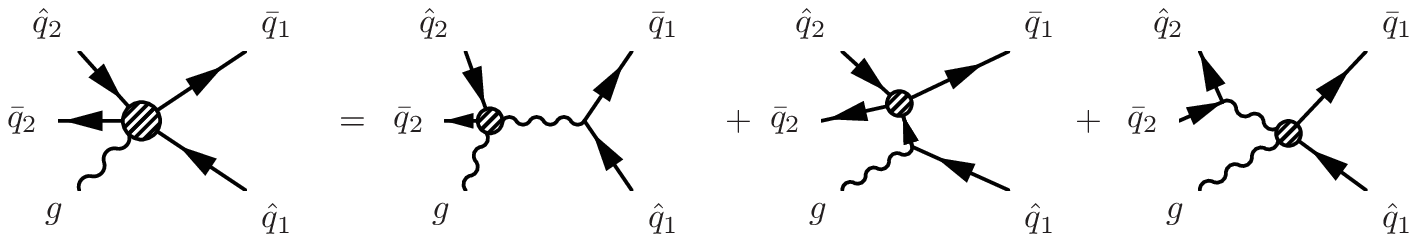}
\caption{Decomposition of the five-point amplitudes after shifting the lines $\hat q_2$ and $\hat q_1$.}
\end{figure}

The amplitude can be written as
\begin{eqnarray}\label{oneGtwoQ}
\mathcal{A}_R( g^{-}, {\bar g_2}^{-1\over 2}, \hat g_2^{-1\over 2}, {\bar g_1}^{-1\over 2}, \hat g_1^z)
&=&\sum_h \mathcal{A}_L( g^-, \bar q_2, \hat q_2, \hat g_{-P_{q_1\bar q_1}}^{-h}){1\over \hat P_{q_1\bar q_1}^2}\mathcal{A}_R(\hat g_{P_{q_1\bar q_1}}^{h},\bar q_1, \hat q_1)\nb\\
&+&\sum_s \mathcal{A}_L(g^-, \hat q_1,{\bar q_1}^{-P_{g\hat q_1},-s}){1\over \hat P_{g\hat q_1}^2-m_1^2}\mathcal{A}_R(\hat q_1^{P_{g\hat q_1},s}, \bar q_2,\hat q_2, \bar q_1)\nb \\
&+&\sum_h \mathcal{A}_L( \bar q_1, \hat q_1, g^-,\hat g_{-P_{\bar q_2\hat q_2,-h}} ){1\over \hat P_{\bar q_2\hat q_2}^2}\mathcal{A}_R(\hat g_{P_{\bar q_2\hat q_2,h}},\bar q_2, \hat q_2)
 \end{eqnarray}
In the spinor form, the amplitude become
\bea\label{AmptwoQz2}
&&\mathcal{A}_R( g^{-}, {\bar g_2}^{-1\over 2}, \hat g_2^{-1\over 2}, {\bar g_1}^{-1\over 2}, \hat g_1^z)=\nb\\
&& {1\over  P_{q_1\bar q_1}^2}{[\td\bt_{\bar q_2},\widetilde{\lmd_2\circ p_{q_2}}]\langle\lmd_I,\lmd_2\rangle\langle\lmd_I,\lmd_{q_2}\rangle +[\td\bt_{\bar q_2},\widetilde{\lmd_I\circ p_{q_2}}]\langle\lmd_2,\lmd_I\rangle\langle\lmd_2,\lmd_{q_2}\rangle\over 4 p_{q_2}\cdot \hat p_{g_I}  p_{g_2}\cdot \hat p_{g_I}}{\langle\lmd_{\bar q_1}, a\rangle[\td b,\td\lmd_I]\over \langle a,\lmd_I\rangle}\nb\\
&+&{1\over  P_{q_1\bar q_1}^2} {[\td\lmd_I,\widetilde{p_{q_2}\circ\lmd_2}]\left(-{1\over m}[ \widetilde{p_{q_2}\circ\lmd_2},\td\bt_{q_2}][\td\bt_{\bar q_2},\td\lmd_I]+[\td\lmd_I,\td\bt_{q_2}]\langle\lmd_{\bar q_2},\lmd_2\rangle\right)\over 4 p_{q_2}\cdot \hat p_{g_I} p_{g_2}\cdot \hat p_{g_I}} {[\td\bt_{\bar q_1},\td b] \langle\lmd_I, a\rangle\over [\td b,\td\lmd_I]} \nb\\
&+& {1\over P^2_{ g \hat q_1}-m_1^2} {2\over (\hat P_{\bar q_2 \hat q_2})^2} {\langle a,\lmd_1\rangle\over [\td\lmd_1,\td b]}\nb\\
&\times&\left(\langle \lmd_{\bar q_2}|\hat P_{g\hat q_1}|\td b][\td \bt_{q_2},\td\bt_{\bar q_1}]-m\langle\lmd_{\bar q_1},\lmd_{\bar q_2}\rangle[\td \bt_{q_2},\td b]+\langle\lmd_{ q_2}|\hat P_{g\hat q_1}|\td b][\td\bt_{\bar q_2},\td\bt_{\bar q_1}]-m\langle\lmd_{\bar q_1},\lmd_{q_2}\rangle[\td\bt_{\bar q_2},\td\bt_{\bar q_1}]\right)\nb \\
&+& {1\over  P_{q_2\bar q_2}^2}{[\td\bt_{\bar q_1},\widetilde{\lmd_I\circ p_{q_1}}]\langle\lmd_1,\lmd_I\rangle\langle\lmd_1,a\rangle +[\td\bt_{\bar q_1},\widetilde{\lmd_1\circ p_{q_1}}]\langle\lmd_I,\lmd_1\rangle\langle\lmd_I,a\rangle\over 4 p_{q_1}\cdot \hat p_{g_1}  p_{g_I}\cdot \hat p_{g_1}}{\langle\lmd_{\bar q_2}, \lmd_{q_2}\rangle[\td\bt_{q_2},\td\lmd_I]\over \langle\lmd_{q_2},\lmd_I\rangle}\nb\\
&+&{1\over  P_{q_2\bar q_2}^2} {[\td\lmd_1,\widetilde{p_{q_1}\circ\lmd_I}]\left(-{1\over m}[ \widetilde{p_{q_1}\circ\lmd_I},\td b][\td\bt_{\bar q_1},\td\lmd_1]+[\td\lmd_1,\td b]\langle\lmd_{\bar q_1},\lmd_I\rangle\right)\over 4 p_{q_1}\cdot \hat p_{g_1} p_{g_I}\cdot \hat p_{g_1}} {[\td\bt_{\bar q_2},\td\bt_{q_2}]\langle\lmd_I, \lmd_{q_2}\rangle\over [\td\bt_{q_2},\td\lmd_I]}
\eea

According to (\ref{AmptwoQz1}) and (\ref{AmptwoQz2}), we can get the amplitude with independent spin states. For example the amplitude with spin ${-1\over 2}$ for massive lines and helicity $-$ for the massless gluon is
\bea
\mathcal{A}_R( g^{-}, {\bar g_2}^{-1\over 2},  g_2^{-1\over 2}, {\bar g_1}^{-1\over 2},  g_1^{-1\over 2}) &=&{\langle\lmd_{q_1},p_{q_1}\circ\td\bt_{q_2}\rangle\over m\langle p_{q_1}\circ\td\bt_{q_2},\lmd_1\rangle }\mathcal{A}_R(\hat g^{-}, {\bar g_2}^{-1\over 2}, g_2^{-1\over 2}, {\bar g_1}^{-1\over 2}, \hat g_1^z)\nb\\
&+&{ \langle\lmd_1,\lmd_{q_1}\rangle\over \langle\lmd_1, p_{q_1}\circ\td\bt_{q_2}\rangle} \mathcal{A}_R( g^{-}, {\bar g_2}^{-1\over 2}, \hat g_2^{-1\over 2}, {\bar g_1}^{-1\over 2}, \hat g_1^z).
\eea
Similarly, the amplitudes with other spin configurations can be obtained by acting with little group generators on them.

\section{Conclusion}
In the article, we discuss the scheme of shifting two massive lines. Moreover we also consider the case that one of the lines is massive and the other is massless. We find it is possible to choose the spin states for the massive line such that the massive fields are $z$-independent under the momentum shift. Meanwhile we can also choose the proper momentum shift scheme such that the massless external fields tend to $1\over z$ under $z\longrightarrow\infty$. According to these properties, we get the recursion relations for the amplitudes which include $\leq 1$ external gluon. We find such amplitude can be divided into two parts which are of relevant spin states between the two shift lines. Each parts of the amplitudes in QCD are constructible directly when we choose proper momentum shift scheme.

The general procedure for the amplitude with fewer than two external massless lines in QCD are as follows. At the beginning, we set a massive line $l_z$ as one of the shifted line in the two line-shift scheme. Another shifted line $l_1$ can be either massive or massless. The spin configuration for other massive line is fixed. The line $l_1$ should be chosen such that the amplitude is constructible for at least one spin state of $l_z$ under a momentum shifting scheme. Then we solve a $z$-independent spin state for $l_z$ and we find it should be related to a spinor in $l_1$ and also dependent on the momentum shifting. According to the BCFW recursion relations, we get the amplitude with this particular spin configuration.

Second, we replace $l_1$ with some other line $l_2$ such that the spin state of $l_z$ of the constructible amplitude are different for the former case. Other spin and helicity configuration is same as before. Similar as the first step, we can get the amplitude with this spin and helicity configuration by the BCFW recursion relations.

Third, we calculate the amplitude with arbitrary spin state for line $l_z$ by linear combination of the two directly constructible amplitudes.  This is a result of the linear property of the amplitude with respect to the external fields.

Fourth, we act spinor forms of the little group generators to the amplitude with a fixed spin configuration for the massive line. We can obtain the amplitudes with any spin configuration directly.

As examples, we calculate the  results of spinor form up to 5-parton amplitudes. And we also compare our results with the usual calculation in Feynman rules. Absolutely they agree with each other exactly.   The amplitudes with more external lines are also easy to obtain. In principle, it is probable to write a computer programm for this procedure to calculate any tree level amplitude in QCD. For the theory beyond QCD, the procedure is also very powerful in the calculation of the amplitudes even though there are two many massive fields in the external fields. Especially for the theory with massive high-spin particles, we will discuss the amplitudes in another independent project.
\section*{Acknowledgement}

This work is supported in parts by NSFC grant No.~10775067  as well as   Research Links Programme of Swedish Research Council under contract No.~348-2008-6049.


\begin{thebibliography}{99}
\bibitem{Parke:1986gb}
  S.~J.~Parke and T.~R.~Taylor,
  {\em An Amplitude for $n$ Gluon Scattering,}
  Phys.\ Rev.\ Lett.\  {\bf 56}, 2459 (1986).

\bibitem{Xu:1986xb}
  Z.~Xu, D.~H.~Zhang and L.~Chang,
  {\em Helicity Amplitudes for Multiple Bremsstrahlung in Massless Nonabelian
  Gauge Theories},
  Nucl.\ Phys.\  B {\bf 291}, 392 (1987).

\bibitem{Berends:1987me}
  F.~A.~Berends and W.~T.~Giele,
  {\em Recursive Calculations for Processes with n Gluons,}
  Nucl.\ Phys.\  B {\bf 306}, 759 (1988).

\bibitem{Dixon1} L. Dixon, {\em Calculating scattering amplitudes efficiently}, Boulder TASI \textbf{95}: 539-584 , [arXiv: hep-ph/9601359].
\bibitem{Dixon2} Z. Bern, L. J. Dixon, D. C. Dunbar and D. A. Kosower, {\em One loop N Point gauge theory amplitudes, Unitarity and collinear limits}, Nucl. Phys. B \textbf{425}, 217 (1994), [arXiv: hep-ph/9403226].
\bibitem{Dixon3} Z. Bern, L. J. Dixon, D. C. Dunbar and D. A. Kosower, {\em Fusing gauge theory tree amplitudes into loop amplitudes}, Nucl. Phys. B \textbf{435}, 59 (1995), [arXiv: hep-ph/9409265].
\bibitem{Dixon4} Z. Bern, V. Del Duca, L. J. Dixon, and D. A. Kosower, {\em All non-maximally-helicity-violating one-loop seven-gluon amplitudes in N=4 super-Yang-Mills thoery}, Phys. Rev. D \textbf{71}, 045006 (2005), [arXiv: hep-th/0410224].

\bibitem{Chalmers} G. Chalmers and W. Siegel, {\em Simplifying algebra in Feynman Graphs Part III: massive vectors}, Phys. Rev. D \textbf{63}, 125027 (2001), [arXiv: hep-th/0101025].
\bibitem{Witten1} E. Witten, {\em Perturbative gauge theory as a string theory in twistor space}, Commun. Math. Phys. \textbf{252}, 189 (2004),  [arXiv: hep-th/0312171].

\bibitem{Britto:2004nj}
  R.~Britto, F.~Cachazo and B.~Feng,
  {\em Computing one-loop amplitudes from the holomorphic anomaly of unitarity
  cuts},
  Phys.\ Rev.\ D {\bf 71}, 025012 (2005)
  [arXiv:hep-th/0410179].

\bibitem{Britto:2004nc}
 R.~Britto, F.~Cachazo and B.~Feng, {\em Generalized unitarity and
 one-loop amplitudes in N = 4 super-Yang-Mills}, Nucl. Phys. B \textbf{725}, 275-305 (2005),
 [arXiv:hep-th/0412103].


\bibitem{Britto:2004ap}
 R.~Britto, F.~Cachazo and B.~Feng, {\em New recursion relations for
 tree amplitudes of gluons}, Nucl. Phys. B \textbf{715}, 499-522 (2005), [arXiv:hep-th/0412308].

\bibitem{Britto:2005dg}
  R.~Britto, B.~Feng, R.~Roiban, M.~Spradlin and A.~Volovich,
  {\em All split helicity tree-level gluon amplitudes}, Phys. Rev. D \textbf{71}, 105017 (2005),
  [arXiv:hep-th/0503198].

\bibitem{Britto:2005fq}
R.~Britto, F.~Cachazo, B.~Feng and E.~Witten, {\em Direct proof of
tree-level recursion relation in Yang-Mills theory}, Phys. Rev. Lett. \textbf{94}, 181602 (2005),
[arXiv:hep-th/0501052].

\bibitem{Britto:2005ha}
  R.~Britto, E.~Buchbinder, F.~Cachazo and B.~Feng,
  {\em One-loop amplitudes of gluons in SQCD}, Phys. Rev. D \textbf{72}, 065012 (2005),
  [arXiv:hep-ph/0503132].

\bibitem{Cachazo:2004dr}
  F.~Cachazo,
  {\em Holomorphic anomaly of unitarity cuts and one-loop gauge theory
  amplitudes},
  [arXiv:hep-th/0410077].

\bibitem{Cachazo:2004kj}
F.~Cachazo, P.~Svrcek and E.~Witten, {\em MHV vertices and tree
amplitudes in gauge theory}, JHEP {\bf 0409}, 006 (2004)
[arXiv:hep-th/0403047].

\bibitem{Cachazo:2004zb}
  F.~Cachazo, P.~Svrcek and E.~Witten,
  {\em Twistor space structure of one-loop amplitudes in gauge theory},
  JHEP {\bf 0410}, 074 (2004)
  [arXiv:hep-th/0406177].

\bibitem{Cachazo:2004by}
  F.~Cachazo, P.~Svrcek and E.~Witten,
  {\em Gauge theory amplitudes in twistor space and holomorphic anomaly},
  JHEP {\bf 0410}, 077 (2004)
  [arXiv:hep-th/0409245].

\bibitem{Cachazo:2005ca}
  F.~Cachazo and P.~Svrcek,
  {\em Tree level recursion relations in general relativity},
  [arXiv:hep-th/0502160].

\bibitem{Badger1} S. D. Badger, E. W. Glover, V. V. Khoze and P. Svrcek, {\em Recursion relations for gauge theory amplitudes with massive particles}, JHEP \textbf{0507}, 025 (2005), [arXiv: hep-th/0504159].

\bibitem{Badger2} S. D. Badger, E. W. Glover, V. V. Khoze and P. Svrcek, {\em Recursion relations for gauge theory amplitudes with massive vector bosons and fermions}, JHEP \textbf{0601} (2006) 066 [arXiv: hep-th/0507161].

\bibitem{Ozeren} K. J. Ozeren and W.  J. Stirling, {\em Scattering amplitudes with massive fermions using BCFW recursion}, Eur. Phys. J. C \textbf{48}, 159-168 (2006), [arXiv: hep-ph/0603071].

\bibitem{Schwinn} C. Schwinn, {\em twistor-inspired construction of massive quark amplitudes}, Phys. Rev. D \textbf{78}, 085030 (2008), [arXiv: 0809.1442].

\bibitem{cohen}
  T. Cohen, H. Elvang, M. Kiermaier
  {\em On-shell constructibility of tree amplitudes in general field theories}, J. High Energy
Phys. \textbf{04},053 (2011), [arXiv: 1010.0257].

\bibitem{Georgiou:2010mf}
  G.~Georgiou and G.~Savvidy,
  {\em Production of non-Abelian tensor gauge bosons: Tree amplitudes in
  generalized Yang-Mills theory and BCFW recursion relation,}
  arXiv:1007.3756 [hep-th].
\bibitem{Cachazo1} P. Benincasa, and F. Cachazo, {\em Consistency conditions on the S-Matrix of massless particles},[arXiv: 0705.4305].

\bibitem{chen}
G. Chen, K. G. Savvidy
{\em Spinor form for the massive fields with half-integral spin}, arXiv:1105.3851.


\end{thebibliography}

\end{document}